\documentclass[useAMS,usenatbib]{mn2e}
\usepackage[T1]{fontenc}
\usepackage{aecompl}
\usepackage{graphicx}

\usepackage{times}
\usepackage{booktabs}
\usepackage[english]{babel}
\usepackage{graphicx}
\usepackage{journals}
\usepackage{color}
\usepackage{amsmath}

\title
{Constraining the PopIII IMF with high-$z$ GRBs}

\author[Ma et al.]{Q. Ma$^{1,2,6}$, U.~Maio$^{3,4}$, B.~Ciardi$^1$, R.~Salvaterra$^5$\\
$^1$ Max-Planck-Institut f\"ur Astrophysik, Karl-Schwarzschild-Stra\ss e 1, D-85748 Garching bei M\"unchen, Germany\\
$^2$ Purple Mountain Observatory, Chinese Academy of Sciences, Nanjing 210008, China\\
$^3$ INAF -- Osservatorio Astronomico di Trieste, via G. Tiepolo 11, I-34131 Trieste, Italy \\
$^4$ Leibniz-Institut f\"ur Astrophysik, An der Sternwarte 16, D-14482 Potsdam, Germany\\
$^5$ INAF, IASF Milano, via E. Bassini 15, I-20133 Milano, Italy \\
$^6$ China University of Chinese Academy of Sciences, Beijing 100049, China
}

\begin{document}

\date{Accepted?? Received??; in original form??}


\pubyear{2016}

\maketitle

\label{firstpage}

\begin{abstract}
We study the possibility to detect and distinguish signatures of enrichment from PopIII stars in observations of PopII GRBs (GRBIIs) at high redshift by using numerical N-body/hydrodynamical simulations including atomic and molecular cooling, star formation and metal spreading from stellar populations with different initial mass functions (IMFs), yields and lifetimes. PopIII and PopII star formation regimes are followed simultaneously and both a top-heavy and a Salpeter-like IMF for pristine PopIII star formation are adopted.
We find that the fraction of GRBIIs hosted in a medium previously enriched by PopIII stars (PopIII-dominated) is model independent.
Typical abundance ratios, such as [Si/O] vs [C/O] and [Fe/C] vs [Si/C], can help to disentangle enrichment from massive and intermediate PopIII stars, while low-mass first stars are degenerate with regular PopII generations.
The properties of galaxies hosting PopIII-dominated GRBIIs are not very sensitive to the particular assumption on the mass of the first stars.
\end{abstract}


\section{Introduction}
\label{sec:intro}
Long Gamma Ray Bursts (GRBs) are believed to arise from the accretion of material onto black holes (BHs) formed after the death of massive stars.
Some of them are so bright that they have been observed up to extremely
high redshift \cite[][]{Salvaterra2009,Tanvir2009,Cucchiara2011}.
In fact, different models \cite[][]{Bromm2002,Salvaterra2007,deSouza2011,Salvaterra2012,Ghirlanda2015,
Elliott2015} consistently predict that $\sim 3$\% of the GRBs detected
by the {\it Swift} satellite should lie at $z>6$.
Among these, some might also be GRBs from the earliest generation (PopIII) of stars
\cite[][]{Suwa2011, Toma2011, MB2014}.
Therefore, high-$z$ GRBs are thought to be a viable probe of the
early Universe besides quasars and galaxies \cite[see][for a recent
review]{Salvaterra2015}.
Indeed, they can pinpoint the  primordial galaxies responsible for the reionization of the intergalactic medium \cite[][]{Salvaterra2013}, providing unique
information about their gas metallicity and dust content
\cite[][]{Campisi2011},
neutral hydrogen fraction
\cite[][]{Nagamine2008},
local intergalactic radiation field
\cite[][]{Inoue2010}
and stellar populations \cite[][]{Ma2015}.
Moreover, they could add constraints to early cosmic magnetic fields
\cite[][]{Takahashi2011},
dark matter nature
\cite[][]{deSouza2013wdm, MV2015}
and primordial non-Gaussianities
\cite[][]{Maio2012ng}.

The detection of a PopIII GRB (heretheafter GRBIII) will
represent a breakthrough for our knowledge of the early phases of
star formation in the Universe. However, with the exception of the highest
redshifts, the GRBIII rate is expected to be much lower than the
rate of PopII GRBs (heretheafter GRBIIs), because of the minor
PopIII contribution to the cosmic star formation rate (SFR) density
\cite[][]{Salvaterra2013}.  For this reason, it will be more likely
to detect signatures of first stars with GRBIIs rather than directly
with GRBIIIs\footnote{Nonetheless, in selected cases, i.e. very massive PopIII stars, the GRBIII rate may be comparable to that of GRBII which form in a medium enriched by PopIII stars \cite[][]{Ma2015}.}.
Indeed, the afterglow of GRBs, expected to be visible
up to very high redshift (e.g. \citealt{CiardiLoeb2000}), carries
information about the surrounding interstellar medium (ISM) through
measurements of the metal abundance ratios.
For example, GRB 130606A at $z=5.91$ has been observed in the full optical and near-IR wavelength region at intermediate spectral resolution with VLT/X-shooter  \citep{Hartoog2015}.
The spectrum has very high S/N ratio with many metal absorption lines detected.
Indirect signatures of the first stars can be identified by exploring the gas enrichment patterns around GRBIIs \cite[][]{Salvaterra2013, Ma2015}.
GRBIIs formed in an environment pre-enriched by massive PopIII stars
(referred to as GRBII$\to$III in \citealt{Ma2015}) are expected to be
preferentially hosted in galaxies with SFR~$\sim 10^{-3}-10^{-1} \, {\rm  M}_{\odot}\,{\rm yr}^{-1}$
and metallicity $Z\sim 10^{-4}-10^{-2}\,{\rm Z}_{\odot}$, lower than those of normal high-$z$ galaxies \cite[][]{deSouza2013,Salvaterra2013}.

Compared to other indirect detection techniques of first stars, e.g. very metal-poor (VMP) and extremely metal-poor (EMP) stars, and also damped Lyman-$\alpha$ absorbers (DLA) \cite[see review e.g.][]{Nomoto2013}, observations of GRBs can extend to very high redshift, where the contribution of first stars is more significant \citep{Ma2015}.
Indeed, GRBs have been observed at $z > 9$ \cite[][]{Cucchiara2011}.
Instead, because of their faint luminosity, VMP and EMP stars are only observed in our Milky Way and nearby dwarf galaxies.
Even so, their metal abundances could be representative of a single (or a few) supernovae episode from first star,
at least for the few cases in which the total metallicity is below the critical value, e.g. the \cite{Caffau2011} star.
In principle, measurements of the abundance ratios in EMP stars could provide a hint about the IMF of PopIII stars.
Also DLAs are observed at high redshift, up to $z \sim 7$ \cite[][]{Becker2012,Simcoe2012}, but their number depends on the density of background quasars, which are very rare at $z>10$.

However, the metal yields from PopIII supernova explosions depend strongly on the properties of their progenitor star
\cite[][]{Heger2002,Heger2010}, and are thus highly uncertain, because
of our persistent ignorance of the typical mass of PopIII stars, which in the current literature is predicted to be both large (e.g. \citealt{Schneider2002, Heger2002, Suda2010}) and small (e.g. \citealt{Clark2011, Stacy2014}).
To improve our knowledge in this respect, here we run numerical hydrodynamical chemistry simulations with different PopIII initial mass functions (IMFs) and study how gas pollution of GRBII hosts is affected by these changes.
The brightness of GRBs which can be observed by {\it Swift}/BAT and also the predicted observable GRB rate relating to first stars have been studied in \cite{Campisi2011} and \cite{Ma2015}, e.g. $\sim 0.06\, \rm yr^{-1}\, sr^{-1}$ of PopIII star enriched GRBIIs should be bright enough to trigger {\it Swift}/BAT.
In this paper we focus on the effect of different first star IMFs on the fraction of GRBIIs triggered in a medium enriched by PopIII stars, and on the properties of GRBII host galaxies.
We will also show how metal abundance ratios detectable in the GRB afterglow spectra of current or future spectroscopic observations can help in discriminating among different PopIII IMFs.

Throughout this work, a standard $\Lambda$CDM cosmological model is adopted with the following parameters:
cosmological constant density parameter $\Omega_{0,\Lambda}=0.7$,
total matter density parameter $\Omega_{0,M}=0.3$,
baryon matter density $\Omega_{b}=0.04$,
primordial spectral index $n=1$,
cosmic variance within a sphere of 8~$\rm kpc/{\it h}$ radius $\sigma_8=0.9$ and
expansion parameter $h=0.7$ in units of 100~${\rm km \, s^{-1} \, Mpc^{-1}}$.
This paper is organised as follows:
the simulations we used are described in Sec.~\ref{sec:simu}, as well as the classification for gas particles;
we present our results in Sec.~\ref{sec:grb};
we critically discuss the caveats of our approach and give our conclusions in Sec.~\ref{sec:conclusions}.

\section{Simulations}
\label{sec:simu}

The code used here is a modified version of GADGET2 code \cite[][]{Springel2005} based on our previous works \cite[see e.g.][for further details]{Maio2007,Tornatore2007,Maio2010,Maio2013} and, besides gravity and hydro, it follows atomic and molecular cooling based on H, He, H$_2$, HD, stellar evolution and metal pollution for various heavy elements (C, N, O, Ne, Mg, Si, S, Ca, Fe, etc.).
To describe physical processes in the interstellar medium which are not directly resolved in the simulation, e.g. star formation in gas particles and feedback from stars, a subgrid model has been adopted. At the end of their lifetimes, which depend on the stellar mass, the stars explode and spread metals and energy into gas neighbours according to the SPH kernel \cite[][]{Tornatore2007}.

We run three simulations with different PopIII IMFs, which are referred to as Very Massive SN (VMSN), Massive SN (MSN) and Regular SN (RSN), as listed in Table~\ref{imf_table}.
The box side length is $10\, {\rm Mpc}\, h^{-1}$ with particles number $2\times 320^{3}$, yielding a gas and dark matter particle mass of $3.39\times 10^{5}\,{\rm M}_{\odot}\, h^{-1}$ and $2.20\times 10^{6}\,{\rm M}_{\odot}\, h^{-1}$, respectively.
All models adopt IMFs with Salpeter slope, but they differ in the lower/upper mass limits and the range of masses contributing to metal pollution.
In the VMSN model, the first stars are assumed to be very massive, in the range [100,~500]~${\rm M_\odot}$, while
the stars contributing to metal spreading are the progenitors of Pair-Instability Supernovae (PISN) in the mass range [140,~260]~$\rm M_\odot$ \cite[][]{Heger2002}.
Both MSN and RSN models have a PopIII IMF covering masses over
[0.1,~100]~$\rm M_\odot$.
In the RSN case, PopIII stars with mass
[40,~100]~${\rm M}_{\odot}$ are assumed to collapse directly into BHs, so that the only contribution to metal enrichment comes from the mass range [10,~40]~${\rm M}_{\odot}$ \cite[][]{Woosley1995, Heger2002}.
In the MSN scenario, instead, stars with masses in the range [40,~100]~${\rm M}_{\odot}$ also contribute to metal pollution by exploding as core-collapse SNe \cite[][]{Heger2010}.
For PopII stars we adopt a Salpeter IMF in the mass range [0.1,~100]~${\rm M}_{\odot}$ in all the simulations.

The various metal yields are tracked during the simulations
and summed up to give the total metallicity $Z$ of each star or gas particle.
The transition from a PopIII to a PopII/I star formation regime is dictated by the local gas metallicity.
More specifically, PopIII (PopII/I) stars are formed at metallicities below (above)
$10^{-4}\, \mathrm{Z}_{\odot}$ \cite[][]{Schneider2002,Schneider2003}.
For PopII stars we include metal yields from AGBs \cite[][]{vandenHoek1997}, type~Ia SNe (SNIa; \citealt{Thielemann_et_al_2003}) and Type~II SNe (SNII; \citealt{Woosley1995}).
\begin{table}
\centering
\begin{tabular}{c c c c}
\hline \hline
Model  & PopIII range [$\rm M_\odot$] & SN range [$\rm M_\odot$] & PopII range [$\rm M_\odot$] \\
\hline
VMSN & $100-500$  & $140-260$ & $0.1-100$ \\
MSN   & $0.1-100$   & $10-100$   & $0.1-100$ \\
RSN   & $0.1-100$   & $10-40$     & $0.1-100$ \\
\hline
\end{tabular}
\caption{
From left to right the columns refer to: model name, stellar mass range for the PopIII IMF, stellar mass range for SN explosions and stellar mass range for the PopII IMF.
}
\label{imf_table}
\end{table}

Metal pollution by PopIII stars is followed separately from that by PopII/I stars.
More specifically, for each gas particle
in the simulated boxes, the fraction of metals produced by PopIII stars is defined as:
\begin{equation}
\label{fiii}
f_{\mathrm{III}} = \frac{\sum_{j} m_{Z_j,{\rm III}}}{\sum_{j} m_{Z_j}},
\end{equation}
where $m_{Z_j}$ is the mass of metal element $Z_j$ in the gas particle, while $m_{Z_j,{\rm III}}$ is the mass coming from PopIII stars.
Here $Z_j$ indicates all heavy elements except from hydrogen and helium.
According to the value of $f_{\mathrm{III}}$, we assign each gas particle with non-zero $Z$ to one of the following three classes:
\begin{itemize}
\item[--] PopII-dominated, if $f_{\rm III}<20\%$;
\item[--] intermediate, if $20\%<f_{\rm III}<60\%$;
\item[--] PopIII-dominated, if $f_{\rm III}>60\%$.
\end{itemize}
We have verified that the exact boundaries chosen for the class definition (e.g. $\pm 10\%$) do not have a relevant impact on the results presented in the paper.

\section{Results}
\label{sec:grb}

In the following, we analyse the simulations and show results of the GRB rate evolution, the probability distribution of different class of
GRBIIs given one or two metal abundance ratios, and also the properties of GRB host galaxies in the three models\footnote{While in principle metal absorption observed in the afterglow of GRBs may be due to either the IGM or the ISM, here we assume that it is representative of the host physical properties.}.
For the metal abundance, we only consider selected elements, such as carbon (C), oxygen (O), silicon (Si)
and iron (Fe), since these are the most abundant in the Universe and easier to detect in the spectra of high-$z$ GRB afterglows \cite[][]{Kawai2006,Castro2013}.
Sulfur (S) always follows Si and shows a similar behaviour.

Quantitatively, metal abundance ratios with respect to the solar values are defined as:
\begin{equation}
{\rm  [A/B]} = \log_{10} (N_{\rm A}/N_{\rm B})  -  \log_{10} (N_{\rm A}/N_{\rm B})_\odot,
\end{equation}
where ${\rm A}$ and ${\rm B}$ are two arbitrary species, $N_{\rm A(B)}$ is the number density of element ${\rm A(B)}$, and the subscript $\odot$ denotes the solar values from \cite{Asplund2009}.

\subsection{GRB rate evolution}
%
At any given redshift, the comoving GRBII rate density in class $i$ sub-sample (i.e. PopII-dominated, intermediate or PopIII-dominated) is calculated as in \cite{Campisi2011} and Ma et al. (2015; hereafter referred to as Ma2015):
\begin{equation}
\label{eq:grb_rho}
\rho_{{\rm GRBII},i}(z)=f_{\rm GRBII}  \,\, \zeta_{\rm BHII} \,\, \rho_{\star,i}(z),
\end{equation}
where $f_{\rm GRBII}$ is the fraction of BHs that ignites a GRBII, $\zeta_{\rm BHII}$ is the fraction of BHs formed per unit of PopII/I stellar mass, and $\rho_{\star,i}$ is the comoving SFR density of class $i$ sub-sample at redshift $z$.
We adopt $f_{\rm GRBII}=0.028$ and $\zeta_{\rm BHII}=0.002$ following \cite{Campisi2011} and Ma2015.
It should be noted that, since GRBs are related to the death of massive stars, in our calculations we include only gas particles with non-zero SFR.

\begin{figure}
\centering
\includegraphics[width=0.99\linewidth]{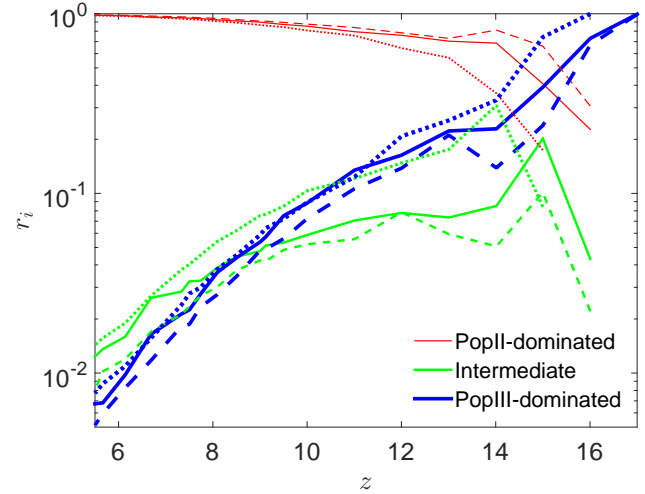}
\caption{
Fraction of GRBII rate which comes from PopII-dominated (thin red lines), intermediate (green) and PopIII-dominated (thick blue) gas as a function of redshift $z$. Solid, dashed and dotted lines denote VMSN, MSN and RSN.
\newline
(The color version is only available in the online journal.)
}
\label{sfrvz}
\end{figure}

In Figure~\ref{sfrvz} we show the redshift evolution of the fraction $r_{i}$ of GRBII rate which is class $i$:
\begin{equation}
r_{i}(z)=\frac{ \rho_{{\rm GRBII},i}(z) }{ \rho_{\rm GRBII,tot} (z) },
\end{equation}
where $\rho_{\rm GRBII,tot}(z)$ is the total comoving GRBII rate density at redshift $z$.
In the figure, solid, dashed and dotted lines refer to VMSN, MSN and RSN respectively.
The three models show a similar evolution, with
the contribution of the PopIII-dominated class decreasing dramatically with redshift, with a fraction of about 1 at $z \gtrsim 17$, but only $\sim 10^{-2}$ at $z=6$.
This is a consequence of the efficient metal enrichment in the early episodes of structure formation.
Indeed, after a first short period in which PopIII events in pristine molecular-driven star forming regions dominate, they rapidly leave room to following generations formed in the recently polluted material.
Interestingly, although the PopIII SFR is somewhat affected by the PopIII stellar properties \cite[][]{Maio2010, Maio2016}, the resulting trends are not very sensitive to the scenarios adopted for the first stars IMF.
The three cases evolve quite closely, with only a minor delay in the RSN scenario, due to the longer stellar lifetimes and later spreading events (consistently with SN ranges in Table~\ref{imf_table}).
The small differences between the trends for VMSN and MSN are due to the different SN explosion energies (lifetimes are comparable), which imply slightly more local enrichment (hence lower PopIII contribution) in the MSN case.
For VMSN metals are spread further away from star forming sites and more diluted within the hosting halo.

The contribution of the intermediate class is always between a few per cents and $\sim 20\%$, with a sharp peak at $z \sim 13-15$ and a subsequent mild drop.
This shape is led by the quick, although transitory, phase of early PopIII enrichment
($f_{\rm III}>60\%$) at $z\gtrsim15$.
The first explosions from short-lived stars rapidly enrich the local medium, and the ongoing spreading events from the newly born PopII stars push $f_{\rm III}$ below the $60\%$ threshold (eq.~\ref{fiii}), causing the steep increase in the figure (green lines).
Due to the increasingly larger amount of metals being expelled from PopII/I stars with decreasing redshift, the values of $f_{\rm III}$ for many star forming gas particles drop below $20\%$. Thus, the contribution associated to the intermediate class
($20\%<f_{\rm III}<60\%$) start to smoothly decrease at $z<15$.

The trend of the PopII-dominated class is complementary to that of the intermediate and PopIII-dominated ones, and gradually kicks in as an increasingly larger fraction of the gas has PopII progenitors due to metal spreading and stellar evolution, i.e. $r_{\rm PopII-dominated}$ increases quickly with decreasing redshift, and it approaches unity for $z \lesssim 6$.

We note that current data at $z\sim 6$ do not show clear signatures of PopIII-dominated GRBIIs (Ma2015) and this is in agreement with the results of Figure~\ref{sfrvz} for $r_{\rm PopIII-dominated}$.

\subsection{Disentangling first stars models}
\label{metal_ab}
%
We discuss here how we can identify PopIII star signals on the basis of the metal abundance ratios measured in the spectra of high-$z$ GRB optical/NIR afterglows.
We also discuss which are the most suitable ratios to look at for constraining the PopIII IMF.
\subsubsection{Single metal abundance ratio}
We start considering the possibility to identify a PopIII-dominated GRBII from  a single metal abundance ratio.
Indeed, the measure of two or more ratios may be challenging for very high-$z$ objects, requiring deep NIR spectroscopy early after the GRB event.

\begin{figure*}
\centering
\includegraphics[width=0.95\linewidth]{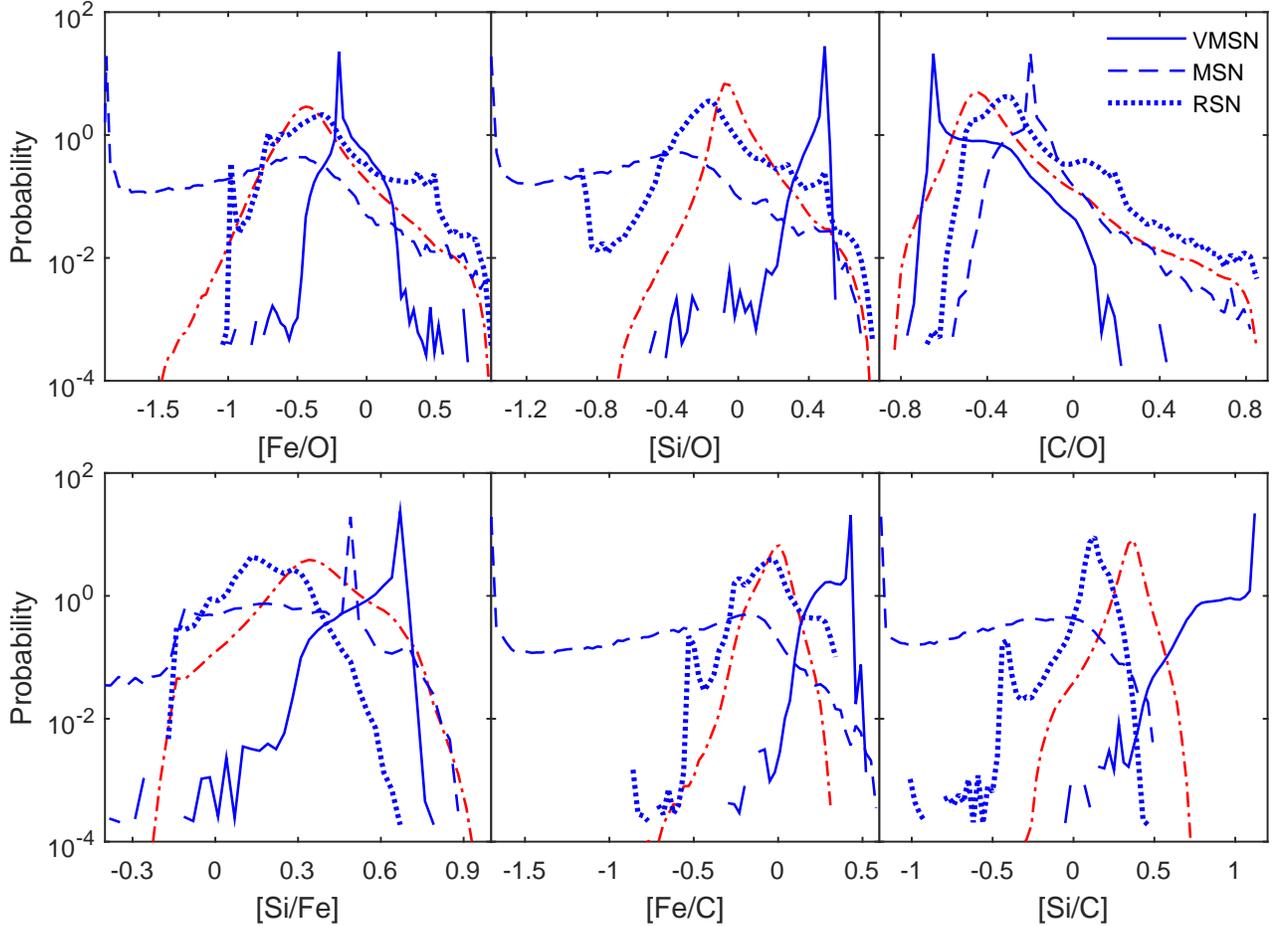}
\caption{
Probability distribution of PopIII-dominated GRBIIs given only one abundance ratio (from left to right and top to bottom, it is [Fe/O], [Si/O], [C/O], [Si/Fe], [Fe/C] and [Si/C]) in model VMSN (solid blue lines), MSN (dashed blue lines) and RSN (dotted blue lines). As a reference, the PDF of PopII-dominated GRBIIs is shown as dash-dotted red line in each panel. Note that the PDFs are normalized by the total rate in each class $i$ and therefore, in number, PopII-dominated GRBIIs overwhelm the distribution of PopIII-dominated ones.
}
\label{pro_dis}
\end{figure*}
The probability distribution of PopIII-dominated GRBIIs at $z \gtrsim
5.5$ as a function of one given abundance ratio is given by:
\begin{equation}
\label{eq:P_1d}
P_{i}(x) =
\frac{1}{\int_{x} \delta \rho_{{\rm GRBII},i}(x)}\frac{\delta \rho_{{\rm GRBII},i}(x)}{\delta x},
\end{equation}
where $x$ denotes one abundance ratio and $\delta \rho_{{\rm GRBII},i}(x)$ is the class $i$ GRBII rate density in the interval
[$x$, $ x + \delta x$].

In Figure~\ref{pro_dis}, we show
the probability distributions of [Fe/O], [Si/O], [C/O], [Si/Fe], [Fe/C] and [Si/C] for PopIII-dominated GRBIIs in the VMSN, MSN and RSN case.
As a reference, we also plot the distribution for PopII-dominated GRBIIs in each panel (the PopII-dominated PDF coincides in the three cases).
We note here that, since the PDFs  are normalized to their total rate (see eq.~\ref{eq:P_1d}), PopII-dominated GRBIIs always overwhelm in number the PopIII-dominated ones.
Therefore, to safely identify PopIII-dominated GRBIIs we should rely only on metal abundance ratio ranges where PopII-dominated GRBIIs are not present.

Let's analyse Figure~\ref{pro_dis} in more detail.
We note that PopII-dominated and PopIII-dominated GRBIIs cover the same range in the [C/O] and [Si/Fe] panels.
Therefore, these abundance ratios are not suitable to disentangle PopIII-dominated GRBs in any of our models.
The VMSN model shows always the smallest dispersion with a sharp peak at
[Fe/O]~$\approx -0.2$,
[Si/O]~$\approx 0.5$
and [C/O]~$\approx -0.6$.
This is consistent with what found by
Ma2015 (see their Fig.~4).
Although the presence of a peak in the probability distribution is a unique feature of the VMSN model, in most cases this can not be identified because it is swamped by normal PopII-dominated GRBIIs.
Therefore, the only unique signature for selecting PopIII-dominated GRBIIs is [Si/C]~$>0.7$, as this characterizes at least 94\% of PopIII-dominated GRBIIs.

The supernova explosions of massive PopIII stars ($> \rm 40~M_{\odot}$) in the MSN model produce very low iron and silicon yields,
but very high carbon and oxygen yields, so that PopIII-dominated GRBIIs in this model peak at very low [Fe/O], [Si/O] and [Fe/C],
and can be identified by e.g.
[Fe/O]~$< -1.5$,
[Si/O]~$< -1$ or
[Fe/C]~$< -0.8$.
Each of these criteria can discern at least 50-60\% of PopIII-dominated GRBIIs. Since the MSN model also includes the contribution from low-mass PopIII stars (with mass $< \rm 40~M_{\odot}$), it has the largest dispersion in each panel.

Finally, PopIII-dominated GRBIIs in the RSN model have a metal abundance distribution very similar to that of PopII-dominated GRBIIs. Although their distributions are visibly shifted compared to those of PopII-dominated
GRBIIs, e.g. in the panel of [Si/O], [C/O] and[Si/C], it would be very difficult to distinguish them from the dominating PopII enriched GRBs.

\subsubsection{Two metal abundance ratios}
Although we have shown that a single metal abundance ratio could be enough to distinguish PopIII-dominated GRBIIs, their identification
would be much easier and more efficient if two or more abundance ratios can be measured.

\begin{figure*}
\centering
\includegraphics[width=0.95\linewidth]{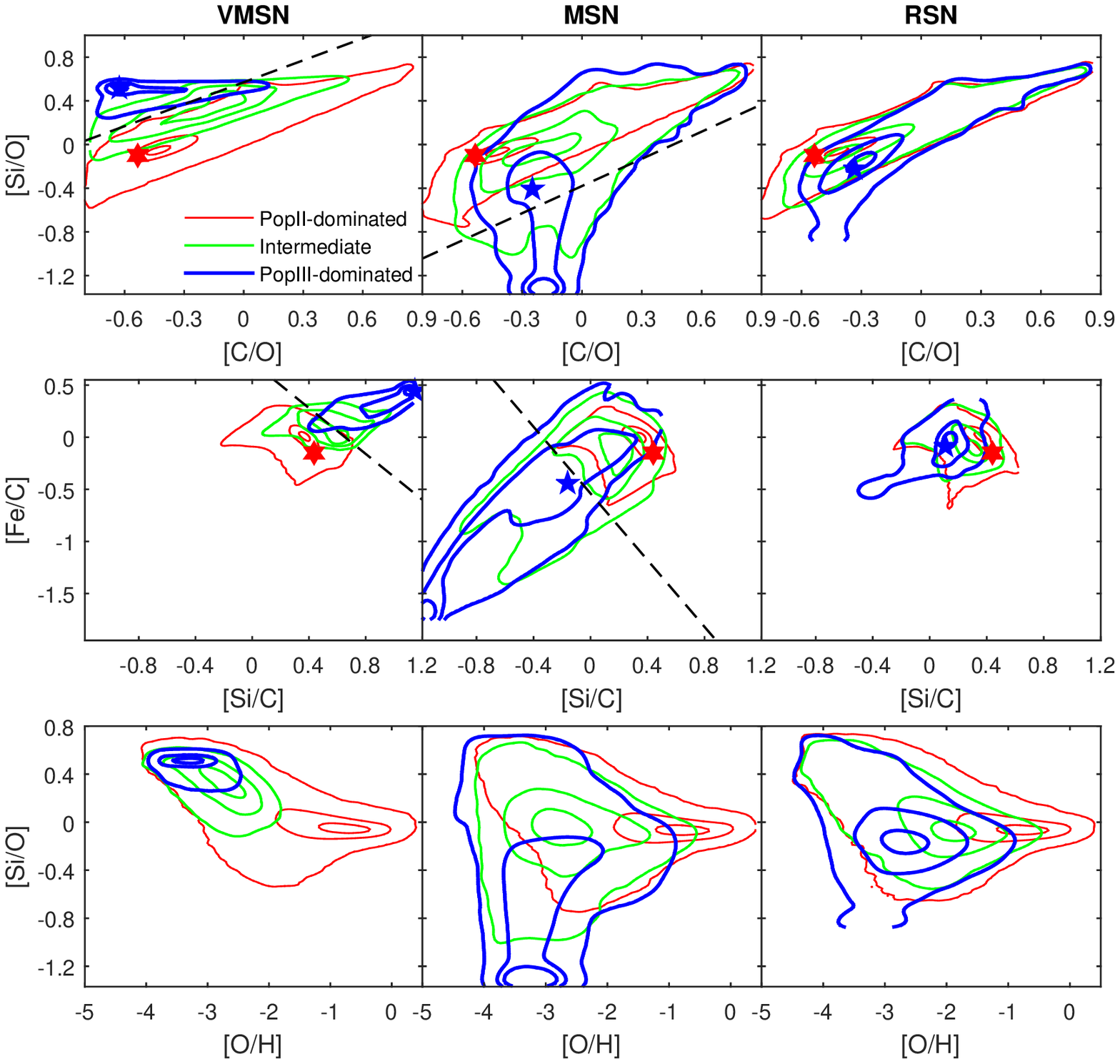}
\caption{Probability of a class $i$ GRBII  to have two given abundance ratios in the plane [Si/O] vs [C/O] (top panels), [Fe/C] vs [Si/C] (middle) and [Si/O] vs [O/H] (bottom), for VMSN, MSN and RSN (from left to right) and PopII-dominated (thin red lines), intermediate (green) and  PopIII-dominated (thick blue) class. The contours of each color refer to a probability of 25\%, 75\% and 100\% from the innermost to the outermost. The black dashed line in the upper left panel corresponds to [Si/O]~$= 0.67 $~[C/O]$ + 0.57$, the one in the upper central panel to
[Si/O]~$ = 0.83 $~[C/O]$ - 0.38$, the one in the middle left panel to
[Fe/C]~$ = -1.1 $~[Si/C]$ + 0.72$, and the one in the middle central panel to [Fe/C]~$ = -1.6 $~[Si/C]$ - 0.54$.
The blue pentagrams and red hexagrams denote the IMF-integrated abundance ratios of the stellar yields from PopIII stars and PopII/I stars respectively.
\newline
(The color version is only available in the online journal.)
}
\label{sisovsco}
\end{figure*}

Figure~\ref{sisovsco} shows the probability density function of hosting a GRBII in environments with given abundance ratios and contribution to metallicity from PopIII stars at redshift $z\gtrsim 5.5$:
\begin{equation}
\label{eq:P}
P_{i}(x,y) =\frac{1}{\int_{x} \int_{y} \delta \rho_{{\rm GRBII},i}(x,y)}\frac{\delta \rho_{{\rm GRBII},i}(x, y)}{\delta x \delta y},
\end{equation}
where ($x,y$) indicate any couple of abundance ratios, while $\delta \rho_{{\rm GRBII},i}(x,y)$ is the class $i$ GRBII rate density in the interval $ \delta x \times \delta y $ centered in $(x,y)$.
Contour levels for probabilities of 25\%, 75\% and 100\% are shown.
Left, central and right columns refer to VMSN, MSN and RSN scenarios.
Upper, middle and lower rows refer to the bivariate probability distributions for
[Si/O] vs [C/O],
[Fe/C] vs [Si/C] and
[Si/O] vs [O/H], respectively.
We note that, since only two metal elements are necessary to plot [Si/O] vs [O/H], the challenge in producing an observational [Si/O] vs [O/H] plot could be the same as that of one single metal abundance ratio, i.e. [Si/O].

Let us consider how different classes of GRBIIs populate the abundance ratio planes for our three simulation runs.
We note that the contours for PopII-dominated GRBIIs look very similar in all models, as their metal signature is only little affected ($<20\%$) by PopIII metal enrichment.
In general, the panel [Si/O] vs [C/O] displays [Si/O] values following a roughly linear relation with [C/O], with a peak probability located around
[C/O]~$ = -0.43$ and
[Si/O]~$ = -0.05$.
The highest probability in the [Fe/C] vs [Si/C] panels is around solar [Fe/C] and slightly supersolar [Si/C].
The [Si/O] ratios evolve as the local metallicity tracked by [O/H] increases, and they converge to [Si/O]~$ \simeq -0.05$ for metallicities [O/H] above $>-2$.

In the VMSN model, PopIII-dominated GRBIIs are located in the upper-left corner of the plane [Si/O] vs [C/O], and can be distinguished from PopII-dominated GRBIIs by selecting those with [Si/O]~$ > 0.67 $~[C/O]$ + 0.57$
(black dashed line in the figure).
The highest probability to identify a PopIII-dominated GRBII is found around
[C/O]~$ \sim -0.62$ and
[Si/O]~$ \sim 0.51$ (almost 75\%),
which is consistent with the critical conditions
[C/O]~$ <- 0.5$ and
[Si/O]~$ >0$
used in Ma2015 to select GRBIIs enriched by PopIII stars.
These can also be identified by selecting
[Fe/C]~$ > -1.1 $~[Si/C]$ + 0.72 $
(black dashed line in the figure),
where the probability peaks at [Si/C]~$ \sim 1.11$ and [Fe/C]~$ \sim 0.43$.
The panel [Si/O] vs [O/H] is not able to discriminate PopIII-dominated
from PopII-dominated GRBIIs, since the distribution of the latter completely encloses the one of the former.

In the MSN model, [Si/O] and [C/O] values of the PopIII-dominated GRBIIs (upper central panel) have a partial overlap with the PopII-dominated class, although they are also located in different regions.
More specifically, at least 70\% of them have [Si/O]~$ < 0.83 $~[C/O]$ - 0.38$, since metal yields from PopIII stars with mass $\rm [40,~100]~M_{\odot}$ have high carbon over oxygen yields, but low silicon and iron \cite[][]{Heger2010}.
The highest probability is around
[C/O]~$ = -0.19$ and
[Si/O]~$ = -1.37$.
The tail distribution of low [Si/C] and [Fe/C] can also be distinguished in the panel [Fe/C] vs [Si/C], with
[Fe/C]~$ < -1.6 $~[Si/C]$ - 0.54$.
The peak values for the PopIII-dominated objects in the [Si/O] vs [O/H] panel have [Si/O]~$ <-1.2$, because the  [Si/O] ratio of the metal yields from PopIII stars in the MSN model is much lower than that from PopII/I stars (which have  [Si/O]~$ \simeq 0$), thus the two classes are completely decoupled in such regimes.

In the RSN scenario, the metal yields of PopIII stars are very similar to those from PopII/I stars.
As a consequence, the contours of PopIII-dominated class overlap with those of other classes, especially the 100\% level, while we can still see some visible, albeit modest, shift for the 25\% and 75\% contour levels
in the same direction as in the MSN case.

Finally, GRBIIs of intermediate class are found to lie between those of the PopII- and PopIII-dominated class in each model.

Irrespectively from the adopted IMF, PopIII-dominated GRBIIs are always found in the low-metallicity range, i.e. at
[O/H]~$ \lesssim -2.0$ (at least for 75\% contours) since O traces total metallicity (roughly speaking, about 2/3 of heavy elements is in oxygen species).
Going to higher [O/H] values, GRBIIs belong to the intermediate- and PopII-dominated class.
This evolution from PopIII- to PopII-dominated class in the different scenarios can be read as a consequence of ongoing mechanical and chemical feedback in the Universe \cite[][]{Maio2011}, which increases the local enrichment level according to the corresponding stellar evolution timescales.

In the panels [Si/O] vs [C/O] and [Fe/C] vs [Si/C], we also show the corresponding IMF-integrated values of the stellar yields from PopIII stars in different models (blue pentagrams) and PopII/I stars (red hexagrams).
While these values could in principle be used to identify signatures of PopIII stars, the distribution of abundance ratios as those obtained with our simulations are much more powerful tools to characterize the yields from different stellar populations.

\subsection{GRB host properties}
\label{host}
%
\begin{figure*}
\centering
\includegraphics[width=0.95\linewidth]{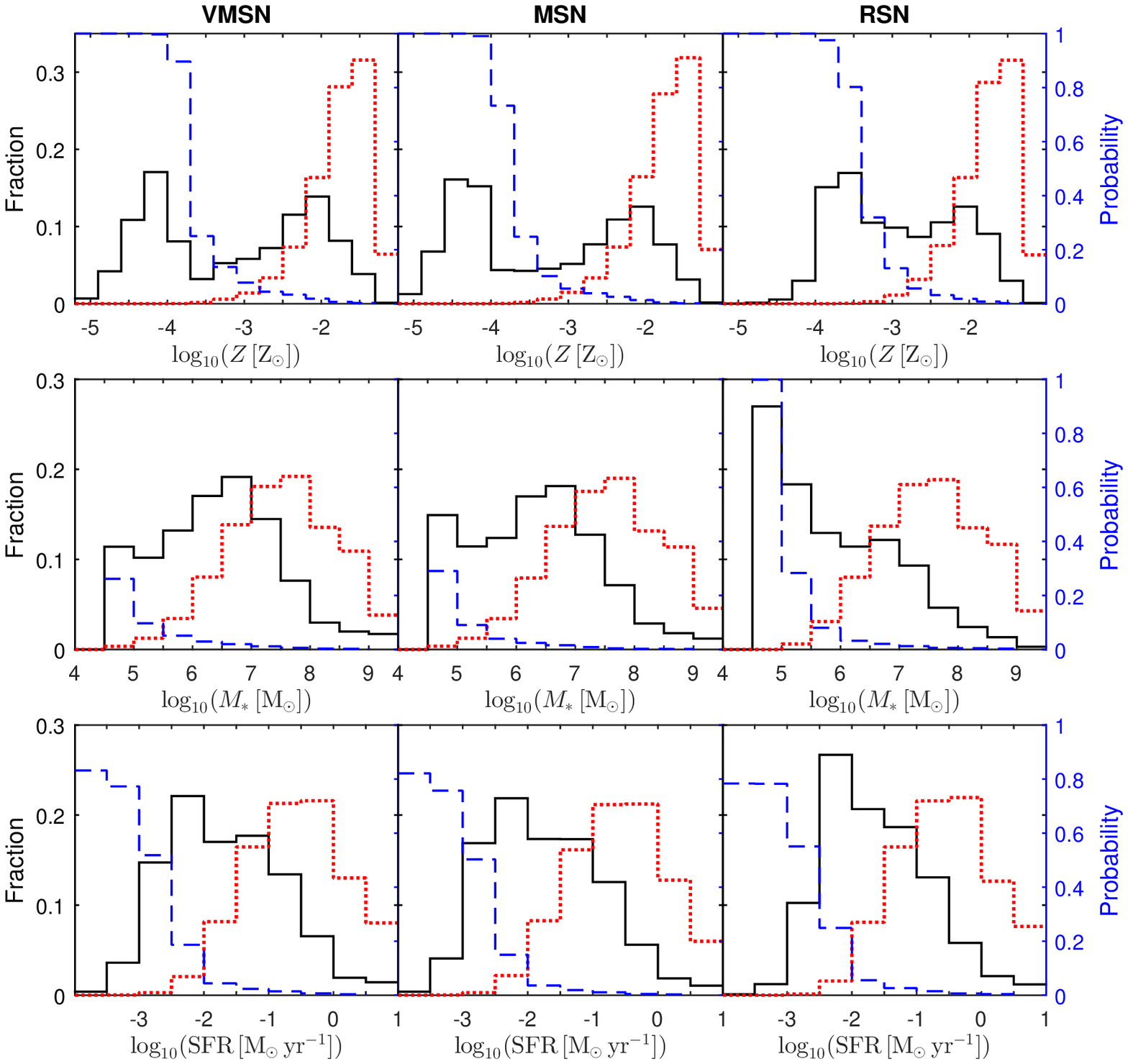}
\caption{From top to bottom, metallicity, stellar mass and SFR distribution of galaxies hosting GRBIIs. The solid black (dotted red) lines are the distributions of galaxies weighted by the PopIII-dominated (PopII-dominated) PopII SFR; while the dashed blue lines refer to the probability for a GRBII hosted in a given galaxy to be PopIII-dominated.
From left to right, the columns refer to model VMSN, MSN and RSN.
}
\label{halo_pro}
\end{figure*}
In Figure~\ref{halo_pro}, we show the PopIII-dominated and PopII-dominated SFR weighted distributions of the properties of GRBII host galaxies in the three models.
We focus on total metallicity $Z$ (top panels), stellar mass $M_\star$ (middle) and SFR (bottom).
As in Ma2015, the weights to the corresponding distributions are computed as SFR$_{i,k}/$SFR$
_{i,{\rm tot}}$, where SFR$_{i,k}$ is the total class $i$ SFR of galaxies in the $k$th bin, and SFR$_{i,{\rm tot}}$ is the integral of class $i$ SFR over all galaxies.
Here we include all the host galaxies at redshift $z\gtrsim 5.5$.
In each panel of the figure, the blue dashed line shows the probability for a GRBII hosted in a given galaxy to be PopIII-dominated, i.e. SFR$_{i,k}$/SFR$_{{\rm tot},k}$, where SFR$_{{\rm tot},k}$ is the total SFR of galaxies in the $k$th bin.

A simple example can better clarify this: let's consider the VMSN model distribution in terms of SFR (bottom-left box).
Although the PopIII-dominated GRBII distribution (solid black line) peaks at SFR~$\sim 10^{-2} \, \rm M_{\odot}\,yr^{-1}$, the probability to find a PopIII-dominated GRBII in a galaxy (dashed blue) with SFR~$\sim 10^{-2} \, \rm M_{\odot}\,yr^{-1}$ is only 20\%.
On the other hand, while a rarer event, the identification of a GRBII in a galaxy with  SFR~$\leq 10^{-3} \, \rm M_{\odot}\,yr^{-1}$ would clearly point towards a PopIII-dominated GRB.

As expected, the different first star models have little effect on the distribution of the properties of PopII-dominated GRBII hosts (dotted red lines),
since the PopII/I regime becomes dominant shortly after the onset of star formation.
More specifically, most of the PopII-dominated GRBIIs are hosted in galaxies with $Z>10^{-2.5} \, \rm Z_{\odot}$, with a peak at $Z=10^{-1.5} \, \rm Z_{\odot}$. Their stellar mass spans the range $(10^{5}-10^{9.5}) \, {\rm M}_{\odot}$, peaking around
$10^{7.5} \, {\rm M}_{\odot}$.
Finally, their SFR is in the range $(10^{-2.5}-10) \, \rm M_{\odot}\,yr^{-1}$, with a peak at $10^{-0.5} \, \rm M_{\odot}\,yr^{-1}$.
These results are consistent with those found by \cite{Salvaterra2013}
and seem to hold for all models.

The metallicity distribution of PopIII-dominated GRBII hosts (solid black lines) shares similar features among the three models.
For example, all of them have a peak at $Z\sim 10^{-2} \, \rm Z_\odot$ and a second one in the range $Z \sim (10^{-4.6}-10^{-3.4}) \, \rm Z_\odot$.
The presence of the $Z \sim 10^{-2} \, \rm Z_\odot$
peak can be explained by residual PopIII star formation in the outskirts of already evolved galaxies (e.g. \citealt{Tornatore2007, Maio2010}). These events can push metallicities close to PopII-dominated ones (red dotted lines).
However, most PopIII-dominated GRBII hosts show metallicities lower than those of PopII-dominated hosts.
In the VMSN and MSN models, the majority of other PopIII-dominated GRBIIs are in galaxies with very low metallicity ($Z \leq 10^{-4} \, \rm Z_\odot$), while in the RSN model the peak of the distribution is at $Z \sim 10^{-3.5} \, \rm Z_\odot$.
This happens as the SFR of PopIII stars in the RSN model is much higher. Although the metal pollution is delayed, a large number of small mass PopIII stars can enrich the galaxies very fast and thus few galaxies have $Z < 10^{-4} \, \rm Z_\odot$.
For all models, the probability to find a PopIII-dominated GRBII (dashed blue lines) becomes lower than 10\% in galaxies with $Z>10^{-3} \, \rm Z_{\odot}$.

The stellar mass of PopIII-dominated GRBII hosts in all models is distributed mainly within the range $(10^{4.5}-10^{7.5}) ~ \rm M_\odot$, lower than that of PopII-dominated GRBII hosts.
However, while in the VMSN and MSN models a peak is present at $M_{\ast} \sim 10^{6.5} ~ {\rm M}_{\odot}$, in the RSN model $\sim$60\% of PopIII-dominated GRBIIs are found in galaxies in the lower stellar mass of the range, i.e. $(10^{4.5}-10^{6}) ~ \rm M_\odot $, with a prominent peak at $10^{5} ~ \rm M_\odot$.
For any stellar mass, the probability of a given galaxy to host PopIII-dominated GRBIIs is very low in the VMSN and MSN model, i.e. PopIII-dominated GRBIIs are not sensitive to the stellar mass of their hosts. In the RSN model, instead, galaxies with $M_{\ast} < 10^{5} \, {\rm M}_\odot$ only host PopIII-dominated GRBIIs.
In fact, metal enrichment from PopIII stars in the RSN model is weaker and delayed (stellar lifetimes are up to 10 times longer than for VMSN), resulting in the survival of more pristine  stars.

The three models also have similar SFR distributions for PopIII-dominated GRBII hosts.
The most populated range is
$(10^{-3}-10^{-0.5}) ~\rm M_\odot\,yr^{-1}$,
with a peak at
$\sim 10^{-2} ~\rm M_\odot\,yr^{-1}$,
about 1~dex lower than that of PopII-dominated GRBII hosts.
In all models, the probability of hosting PopIII-dominated GRBIIs goes from 80\%
for galaxies with
SFR~$\rm < 10^{-3} ~ M_\odot\,yr^{-1}$
to 5\% for those with
SFR~$\rm > 10^{-2} ~ M_\odot\,yr^{-1}$,
with only a few PopIII-dominated GRBIIs in
galaxies with SFR~$\rm > 1 ~ M_\odot\,yr^{-1}$.

\section{Discussion and Conclusions}
\label{sec:conclusions}

In this work we have studied the possibility to distinguish first star models with metal abundance ratios measured in high-$z$ GRB afterglow observations using N-body hydrodynamical cosmological simulations.
The numerical simulations include a detailed chemical evolution, gas cooling and metallicity dependent star formation.
We run three simulations, which differ only for the models for first stars (and thus the associated metal enrichment):
({\it i}) a Very Massive SN (VMSN) model with mass range [100, 500]~$\rm M_{\odot}$ in which metal pollution is driven by PISN with [140, 260]~$\rm M_{\odot}$;
({\it ii}) a Massive SN (MSN) model with mass range [0.1, 100]~$\rm M_{\odot}$ where metals are produced by SNe with [10, 100]~$\rm M_{\odot}$;
and ({\it iii}) a Regular SN (RSN) model which is the same as the MSN model, but here the enrichment is done by SNe with masses of [10, 40]~$\rm M_{\odot}$.

Our calculations are based on a few implicit assumptions that deserve some critical discussion as, in principle, they can affect the final results.
In particular, from the point of view of the stellar properties, the PopIII SFR may be slightly changed by the particular critical metallicity adopted in the range $\sim (10^{-6}-10^{-3}) ~ \rm Z_{\odot}$, without altering the overall cosmic SFR and the PopII/I SFR.
In addition, different stellar structure models may shift the resulting abundance ratios, although no major changes are expected, especially at low metallicities \cite[see discussion in e.g.][]{MT2015}.
In fact, a number of physical processes in stellar cores (differential rotation, the initial composition, magnetic fields, nuclear reaction rates, explosion mechanisms, etc.) may affect theoretical metal  yields, although their specific values are not expected to change significantly cosmic gas evolution
the fraction of GRBIIs in a PopIII-dominated medium, and the properties of GRBII host galaxies.
Indeed, and mostly at early times, metallicities are dominated by oxygen, for which fairly solid constraints exist.
Differently, the abundance ratios of PopIII-dominated GRBIIs
are sensitive to the predicted metal yields from PopIII stars.
For example, massive first stars (i.e. with mass $> 40\,\rm M_{\odot}$ in the MSN model) can explode both as faint supernovae and hypernovae, depending on the angular momentum of the resulting black hole \citep{Nomoto2013}.
The former case produces small heavy element yields (e.g. iron; \citealt[][]{Heger2010,Nomoto2013}), which could explain the abundance ratios of carbon-enriched metal poor (CEMP) stars observed in our Milky Way. In this case, PopIII-dominated GRBIIs in our MSN model can be identified with the abundance ratios and used to study first stars.
The latter case has higher silicon and iron yields, and could be the reason for carbon-unenhanced metal poor stars \citep{Umeda2005}. Their abundance ratios are very similar to those of PopII/I stars, thus it would be difficult to separate those PopIII-dominated GRBIIs from the PopII-dominated ones.
Finally, a different PopIII IMF slope may have an effect on the abundance ratio distributions, especially in the MSN model,
by shifting the peak distribution discussed in Section~\ref{metal_ab}.
However, the ranges of abundance ratios which we focus on are only mildly affected, as they are mainly driven by the adopted SN mass range.

We should additionally note that physical processes in the interstellar medium are described by means of subgrid models, which have also been tested through simulations of isolated objects \cite[e.g.][]{Maio2013iso}.
The effects of changing specific model parameters have been shown to be mild for star formation and enrichment, as long as the gas cooling is properly accounted for in both atomic and molecular phases, as in our simulations.

Most of our results are weighted by the GRBII rates in the gas particles of the corresponding classes.
We have considered that most PopIII stars are located either in the outskirts of massive galaxies \cite[][]{TFS2007} or in low-mass galaxies \cite[][]{Maio2011,Wise2012,BM2013,Maio2016}, i.e. further away from the typical location of PopII stars. This means that GRBIIs exploding in these environments are more likely to retain information from PopIII rather than from PopII/I stars.
A weighting by other properties, e.g. the gas or the metal mass of the particles, would give results very similar to those we have presented, especially concerning the criteria to identify the first star signal.

Theoretically, the amplitude of metal absorption lines depends on the GRB afterglow luminosity, the spectral shape and also the ionization status of the interstellar medium.
To investigate this in more details, one would need to model the luminosity and spectrum of the afterglow \cite[e.g.][]{CiardiLoeb2000}, as well as the ionization status and dust fraction of the interstellar gas at very high resolution.
This is beyond the scope of this paper.
Compared to recent afterglow observations of high-$z$ GRBs, e.g. GRB 111008A \cite[][]{Sparre2014} and GRB 130606A \cite[][]{Hartoog2015}, we find that most of the absorption lines of carbon, oxygen, silicon and iron (except several Fe lines) would be observable even at $z>10$ with VLT/X-shooter.

We highlight that we have defined the dominance of a population according to the fraction of heavy elements coming from PopIII stars, $f_{\rm III}$, with PopIII- and PopII-dominated regimes having $f_{\rm III} > 60\%$ and $f_{\rm III} < 20\%$, respectively.
Nevertheless, we have verified that the exact boundaries used for our classification (within $\pm 10\%$ variations) do not have a relevant impact on the final outcome.

We can summarize our main results as follows.
\begin{itemize}
\item The fraction of second generation GRBs exploded in a medium enriched by PopIII stars (PopIII-dominated GRBIIs) is independent from the adopted first stars model.
This fraction rapidly decreases with redshift,
accounting for $\sim30$\% of GRBII at $z=15$, 10\% at $z=10$, but only 1\% at $z=6$.
Since the PopII SFR is almost independent from the mass spectrum of PopIII stars, also the observable rate of PopIII-dominated GRBIIs is hardly affected by it.
\item We have explored the possibility to identify PopIII-dominated GRBIIs from  observations of abundance ratios.
Although not very efficient, a single ratio could in principle be enough to select PopIII-dominated GRBIIs in the VMSN and MSN model.
In the former case, a unique signature is represented by a high [Si/C] ratio, i.e. [Si/C]~$>0.7$. In the latter, PopIII-dominated GRBIIs can be identified by
[Fe/O]~$ < -1.6$,
[Si/O]~$ < -1 $, or
[Fe/C]~$ < -0.8$.
\item PopIII-dominated GRBIIs are more easily selected using two abundance ratios.
For massive or very massive SN models, the
probability distribution shows memory of the adopted IMF, allowing to distinguish the two.
In the VMSN model, PopIII-dominated GRBIIs are found at
[Si/O]~$ > 0.67 $~[C/O]$ + 0.57$ or
[Fe/C]~$ > -1.1 $~[Si/C]$ + 0.72$, whereas in the MSN model, at least 70\% of them are within the limits of
[Si/O]~$ < 0.83 $~[C/O]$ - 0.38$ or
[Fe/C]~$ < -1.6 $~[Si/C]$ - 0.54$.
On the other hand, as metal yields from PopIII SN explosions in the RSN model are very similar to those of PopII/I SNe, in this case it is not feasible to distinguish PopIII- from PopII-dominated GRBIIs.
\item The properties of galaxies hosting PopIII-dominated GRBIIs are not strongly affected by the assumption on the mass of the first stars.
Generally, these galaxies have metallicity $(10^{-4.5}-10^{-1}) ~ \rm Z_{\odot}$, stellar mass $(10^{4.5}-10^{7.5}) ~ \rm M_{\odot}$ and SFR~$(10^{-3}-10^{-0.5}) ~ \rm M_{\odot} \, yr^{-1}$, all lower than those of the normal PopII-dominated GRBII hosts (in agreement with Ma2015).
Additionally, the GRBIIs observed in host galaxies with $Z<10^{-4}\,\rm Z_{\odot}$ are most likely PopIII-dominated.
\item Finally, we would like to mention that none of the GRBs detected so far at $z=5-6$ shows abundance ratios favouring a PopIII-dominated
environment (see Ma2015 for a case by case discussion).
This is consistent with the expected fraction of PopIII-dominated  GRBIIs in this redshift range, i.e. $\sim 10^{-2}$.
\end{itemize}

\section*{Acknowledgments}
We acknowledge the anonymous referee for careful reading and very helpful comments that improving the presentation of the results.
Q. Ma is also partially supported by the National Natural Science Foundation of China (Grant No. 11373068 and No. 11322328), the National Basic Research Program ("973" Program) of China (Grants No. 2014CB845800 and No. 2013CB834900), and the Strategic Priority Research Program "The Emergence of Cosmological Structures" (Grant No. XDB09000000) of the Chinese Academy of Sciences.
We used the tools offered by the NASA Astrophysics Data Systems  and by the JSTOR archive for bibliographic research.

\bibliography{ref}
\appendix
\end{document}